\documentclass[12pt]{article}

\usepackage{graphicx}% Include figure files

\newcommand{\be}{\begin{equation}}
\newcommand{\ee}{\end{equation}}
\newcommand{\bea}{\begin{eqnarray}}
\newcommand{\eea}{\end{eqnarray}}

\def\k{\kappa}

\begin{document}

\begin{titlepage}

%\begin{flushright}
%{\tt
%    hep-th/yymmnnn}
% \end{flushright}

\bigskip%\vskip3mm

\begin{center}

{\bf{ Hawking radiation from
extremal %versus
and non-extremal black holes}}

\bigskip
\bigskip\bigskip
R. Balbinot$^a$,\footnote{balbinot@bo.infn.it} A.
Fabbri$^b$,\footnote{afabbri@ific.uv.es} S.
Farese$^b$,\footnote{Sara.Farese@uv.es} R.
Parentani$^c$,\footnote{parenta@th.u-psud.fr}

\end{center}

\bigskip%

\footnotesize \noindent {\it a) Dipartimento di Fisica
dell'Universit\`a di Bologna and INFN sezione di Bologna, 40126
Bologna, Italy}\\
{\it b) Departamento de F\'{\i}sica Te\'orica and
    IFIC, Centro Mixto Universidad de Valencia-CSIC.
    Facultad de F\'{\i}sica, Universidad de Valencia,
       46100,Burjassot, Valencia, Spain}\\
{\it c)   Laboratoire de Physique Th\'eorique, CNRS UMR 8627, B\^at. 210,\\ Universit\'e Paris XI, 91405 Orsay Cedex, France}
%non ce niente da vedere}

\bigskip%\vskip2mm

\bigskip%\vskip2mm

\begin{center}
{\bf Abstract}\\
\end{center}
The relationship between Hawking radiation emitted
%R particle creation in
by non extremal and extremal Reissner Nordstr\"om black holes %(BH)
is critically analyzed. A careful study of % analysis of %R
a series of regular collapsing geometries
reveals that the % two-dimensional
stress energy tensor stays regular in the extremal limit
%is regular on their horizon
%extremal BHs are indeed regular
and is smoothly connected to that of non extremal black holes.
The unexpected feature is that the late time transients which played little
role in the non extremal case are necessary to preserve
the well defined character of the flux in the extremal case.
The known singular behavior of the static energy density
% stress energy
%R usually found in
of extremal black holes %ones %BHs
is recovered from our series by neglecting these transients,
%of
%an incorrect
when performing what turns out to be an illegitimate late time limit.
Although our results are derived in two dimensional settings,
we explain why they should also apply to %RS four
higher dimensional black holes.
%leading from from non extremal to extremal.
%configurations.

\end{titlepage}
%%%%%%%%%%%%%%%%%%%%%%%%%%%%%%%%%%%%%%
\setcounter{equation}{0}

\section{Introduction}

%Older text The %Hawking temperature
% surface gravity of charged Reissner-Nordstr\"om black holes
% is given by
% \be
% %T_H
% \kappa(M,Q) =
% \ee
% where $M$ and $Q$, their mass and their charge respectively,
% % of the hole
% obey $M \geq |Q|$.
Non extremal Reissner-Nordstr\"om black holes (NEBH) form a two parameter
family in which the inequality $M > |Q|$ holds,
where $M$ and $Q$ are their mass and their charge respectively.
Extreme Reissner-Nordstr\"om %(RN in the following)
black holes (EBH) form a one parameter family, and obey $M=|Q|$. Their surface gravity vanishes and so does
their Hawking temperature.
%$T_H = \kappa/2\pi$ (in units where $\hbar =1$).
%Extreme Reissner-Nordstr\"om (RN in the following) black holes (BH)
%are charged BHs characterized by the relation between their
%mass $M$ and charge $|Q|$. This implies a vanishing Hawking
%temperature $T_H$, so they
Therefore, if
discharge does not occur, EBH can be regarded as the end point configuration of
the evaporation of %a non-extremal black holes
NEBH.
% (i.e. $M>|Q|$).

However EBH and NEBH  seem rather disjoint
%these two families of BHs, namely extremal and non-extremal,
in many aspects.
 On one hand, the Euclidean section of
EBH is very different than that of NEBH.
%looks quite different from the one associated to a non-extremal BH,
It possesses an infinite throat, the horizon sitting
at the end of it.
Since the geometry is regular when approaching the throat,
%there requiring, $\kappa$,
%periodic identification of the Euclidean time
%variable.
the period of Euclidean time is arbitrary,
unlike for NEBH where the period must be $2 \pi/\kappa$,
where $\kappa$ is their surface gravity,
in order for not having a conical singularity on the horizon. %\\
This has led some authors \cite{Hawking:1994ii, Teitelboim:1994az,
Ghosh:1994mm, Das:1996rn, Hod:2000pa} to %ROne has
conjecture that the Bekenstein-Hawking area-entropy relation does
not apply to EBH which should be characterized by zero entropy.
Since then, this conjecture has been
invalidated by %contradicts microstates counting in
string theory which %R does
confirmed the Bekenstein-Hawking formula by counting microstates of some
particular EBH \cite{Strominger:1996sh}.

On the other hand, using
the more %R most
familiar settings of Quantum Field Theory %(QFT)
in curved space, EBH seem to be plagued by divergences of
%the expectation value of
the stress tensor which are
%R for quantized matter fields
absent for NEBH, thereby reinforcing the idea that EBH should
be considered as forming a disconnected family.
%  %in the non-extremal case.
With more details, when considering the formation of a NEBH by
gravitational collapse, any regular state evolves at late time to
a {\it stationary} state, often referred to as the ``Unruh"
vacuum, which is characterized by the condition of no incoming
flux and by the regularity on the future horizon (as seen in a
freely falling frame). This regularity is ensured by the steady
thermal radiation with
 Hawking temperature $\kappa/2\pi$.
%%%%%R I think we should not mentioned Boulw and HH
%three {\it stationary} quantum states can be unambiguously identified,
%  namely, the "Boulware" vacuum by the condition that the asymptotic fluxes vanish,
% the "Unruh" vaccum by the condition of no incoming flux and regularity on the
% future horizon, and the "Hartle-Hawking" state by regularity on both horizons.
When taking the extremal limit of this stationary situation, that is $ M\to|Q|$,
%the vanishing of the surface gravity
% indicates
% %has lead to the conclusion
% %it has been %R often  stated
% that these three states  coincide.
%in an extremal BH
%and that m Moreover
the outgoing flux disappears since the surface gravity vanishes,
but the resulting stress tensor is found to be singular
\cite{Trivedi:1992vh, Frolov:1987gw, Anderson:1994hg}, in a way
similar to what is obtained in
%(when seen in a freely falling frame), a singular static infinite vacuum energy density,
%polarization,
 the ``Boulware" vacuum of NEBH which is
 the stationary state with no Hawking radiation.
% It should be noticed that these results
% are obtained starting from the relevant quantity
% (expectation value of the stress tensor)
% calculated for %non-extremal
% stationary NEBH and then taking the extremal limit %R letting smoothly
% $ M\to|Q|$.

However it has been also shown \cite{Balbinot:2004jx, Fagnocchi:2005uk} that when one considers directly
the %time-dependent
formation of an EBH % extremal BH
(for example by the collapse of a shell
% of extremely charged matter, i.e.
with $M=|Q|$), the resulting stress tensor
is regular on the future horizon, in agreement with the general analysis of
\cite{Fulling:1978ht}. %, as it is the case for NEBH. %\\
This result seems to contradict %is in striking contradiction with
what we just obtained by % would obtain in
considering first the formation of a NEBH %($M>|Q|$)
and then taking the extremal limit %($M\to|Q|$)
since a singular
stress tensor was found.  % \\
Were this contradiction to persist,
%were the case,
this would establish
%This seems to reinforce
the fact %believe
that NEBH and EBH
%non-extremal and extremal BH
are indeed quantum mechanically %quite
distinct objects since one could not
%smoothly connect
obtain expectation values for EBH from those evaluated with NEBH.
Such conclusion was reached in \cite{Liberati:2000sq} where
%R3 one can read
it is claimed that the extremal case ''in no sense
represents a limit of the nonextremal case but implies a real discontinuity''.
%R3
%``the extremal wordline (trajectory) in no sense
%represents a limit of the nonextremal case but implies a real
%discontinuity in the asymptotic behavior of the collapsing object''.
%one configuration to the other.\\

The purpose of this paper is to show that this conclusion is not
correct.
%study more closely the relation between extremal and non-extremal BHs and show that
% we shall see that expectation values
%smoothly evolve from their NEBH values to their EBH ones.
By a careful analysis of the extremal limit, we shall demonstrate
the {\it continuity} %R I loke this term because it is the opposite of discontinuous
%whereas smooth is rather vague
of the expectation values in %when taking
the limit $M \to |Q|$.
To have a well defined limiting procedure, we shall consider
a series of {\it regular} collapsing geometries with  $M \to |Q|$,
and compute the local fluxes for every value of $(M,Q)$.
Two subtle points are encountered in this limit.
%R3 The first is that (too many that)
Firstly,  the regularity is preserved in the extremal
only by taking into account the late time transients which played
no significant role for NEBH. Moreover the properties of these transients
are independent of the collapse.
%R3 The second point is that
Secondly, the late time limit (giving rise to
stationary fluxes) cannot be taken before
the extremal limit. This demonstrates that {\it stationarity} cannot be assumed
when analysing EBH, at least when dealing with regular collapsing geometries,
as opposed to %RS ill-defined
singular (and hence ill-defined)
eternal configurations.

%is taken allows to solve the above puzzle. %\\

The plan of the paper is the following.
In section 2 we compute %analyze
the stress tensor %Hawking radiation for the formation of a
resulting from the collapse to a NEBH.
%non-extremal BH in terms of expectation value of the associated stress tensor.
The material presented in this section is rather %quite
standard, but with a special attention on transients
in order to prepare taking the extremal limit. %
% However its discussion is necessary since
%only a rigorous comprehension of it will be used
%identify the loophole of the usual approach and find the correct way
 % should be taken.
In section 3 the analysis is repeated with the formation of an extremal BH
and the necessity of keeping the transients is established.
%is shown how a
Then the smooth connection with the results of the previous section is
demonstrated.
% indeed exists.
 Section 4 contains the conclusions. %\\
%To allow analytically manageable calculation to be performed we
%restrict o
Throughout the paper we shall work with
%Our analysis will be done using the
a two-dimensional analytical treatment, and
%to a two dimensional spacetime.
at the end of the paper we argue that our results
should also apply to four dimensional BH.
% applicability
%However the
%final results are expected to hold in higher dimensions also.\\
An appendix %at the end
contains
%the basic formulae for
%the computations of expectation values of the quantum
the expressions of the stress tensor of 2D massless fields
% for minimally coupled massless scalar fields propagating in a 2D spacetime. %
we use in the text.

\section{Hawking radiation emitted by NEBH}
%for charged non-extremal BH}
As shown by Hawking \cite{Hawking:1974sw}, the formation of a BH triggers a vacuum
instability resulting in the %R creation
emission of particles radiated towards
infinity.
%R A
When the BH is non extremal, at late time
and independently on the details of the collapse (besides its
 regularity \cite{Grove:1990gc}),
%there is
one obtains a stationary flow of thermal radiation with Hawking
temperature $T_H=\k/2\pi$, where $\k$ is the surface gravity
of the outer horizon. %\\
The key properties of the associated stress tensor
% major features of this process
can be thus obtained %R visualized in
by considering
%a simple spherically symetric %2D model of the collapse, namely
the formation of a BH by the collapse
of a spherically symetric ingoing null thin shell.
Indeed, the key property to get Hawking radiation is the
regularity of the geometry which is guaranteed when the infalling trajectory
is inertial. As shown in \cite{Birrell:1982ix}, it suffices that the
trajectory be non-singular across the future horizon.

In this Section, %ere we shall
we consider the non extremal case, i.e. $M>|Q|$.
%case where the resulting BH is
%charged but
% The Carter-Penrose diagram corresponding to
%this collapse is depicted in Fig. \ref{Mink_RN_NE}.
 %R Then of some
We consider the collapse of an ingoing charged
null shell located at $v=v_0$.
% generates a non-extremal Reissner-Nordstr\"om BH.
For $v<v_0$ the
spacetime is %flat
Minkowski and the %spacetimes
metric reads (dropping the angular variables)
%taking $\theta,\phi=const.$)
\be ds^2=-du_{in}\, dv,\ee where \bea
u_{in}=t_{in}-r&,&v=t_{in}+r.\label{nullMink}\eea
%R I think that the inside and outside t MUST be different, CORRECT ?
%THIS HAS NO CONSEQUENCE IN WHAT FOLLOWS, THOUGH.
%while f
Outside the shell, for $v>v_0$, one has
 \be\label{NEle}
ds^2=-f(r)dudv=-\bigg(1-\frac{2M}{r}+\frac{Q^2}{r^2}\bigg)dudv,\ee where now
\bea u=t-r^\ast&,&v=t+r^\ast,\label{nullBH}\eea $r^\ast$ being
the tortoise radial coordinate \bea
r^\ast(r;M,Q) %R this "strange" notation is to prepare the lomit M to Q at fixed r
&=&\int^r\frac{dr'}{1-\frac{2M}{r'}+\frac{Q^2}{r'^2}},
\nonumber \\
&=& r + \frac{1}{2\k_+}\ln\left[\k_+(r- r_+)\right]
%\\&&
- \frac{1}{2\k_-}\ln\left[\k_-(r-r_-)\right],
\label{NErstar}
\eea
%R is thi scorrect ?? with the 2s ??
where $\k_\pm$ and $r_\pm$ are the surface gravities
and the radii of the two horizons (outer and inner respectively)
\bea\label{kpm} \k_\pm&=&\frac{\sqrt{M^2-Q^2}}{r_\pm^2},
\nonumber \\
r_\pm &=& M\pm \sqrt{M^2-Q^2}
.\eea
%R if some reader want to check the eqs he will need r_\pm
%R I also consider that these notions shoudl be given before the pasting u(U)
%\begin{figure}
%\begin{center}
%\includegraphics[angle=0,width=1.5in,clip]{Mink_RN_NE.eps}
%\caption{Penrose diagram describing the formation of a non-extremal
%Reissner-Nordstr\"om black hole from Minkowski spacetime by the
%collapse of a null shell.} \label{Mink_RN_NE}
%\end{center}
%\end{figure}
Asymptotic flatness %R ($\mathcal{I}^-$ exists)
implies that the
%R Eddington Finkelstein
ingoing null coordinate $v$ is the same on
both sides of the shell. On the other hand the relation between
$u_{in}$ and $u$ can be found by requiring the continuity of the radial
coordinate $r$ along the shell. From Eqs. (\ref{nullMink}) and
(\ref{nullBH}) evaluated on the shell we have \bea\label{rstarshell}
\frac{v_0-u_{in}}{2}=r&,&\frac{v_0-u}{2}=r^\ast.\eea
Using Eq. (\ref{NErstar}) and eliminating $r$ between the above two equations
 we exactly %R
 get
%have the relation between $u_{in}$ and $u$, namely
\bea\label{uuinne}
u=u_{in}-\frac{1}{\k_+}\ln\left[\k_+(v_0-u_{in}-2r_+)\right]
%\\&&
+\frac{1}{\k_-}\ln\left[\k_-(v_0-u_{in}-2r_-)\right].\eea
From this we see that the event horizon, defined by $u=+\infty$,
%outer one $r=r_+$ is the . It
corresponds to $u_{in}=v_0-2r_+$ and to $r=r_+$.
To simplify the forthcoming equation, we introduce
a new null coordinate \be
U_{in} = u_{in} - v_0 + 2r_+,
\label{newU}
\ee
which vanishes on the event horizon and which is linearly
related to $u_{in}$.
We also notice that in the late
time limit, $u\to\infty$,
%and replacing by a regular shift $u_{in}-2r_+ +v_0$ by $U$, %R
 Eq. (\ref{uuinne}) yields
 \be \label{KruskalUin}
u=-\frac{1}{\k_+}\ln\left(-\k_+ U_{in}\right) + D + O(U_{in}) ,\ee %R CORRECT
 where $D$ is a constant which plays no role as it can be absorbed in $u$.
 When ignoring the linear correction $O(U_{in})$
 %and the irrelevant question
 we recover the usual relation between the
 Kruskal coordinate $U_K$ and the asymptotic coordinate $u$:
  \be \label{KruskalUin2}
u=-\frac{1}{\k_+}\ln\left(-\k_+ U_K\right).\ee This relation could
be obtained by considering the eternal BH geometry, i.e. without
referring to any collapse. As we shall see, the important physical
consequence of the late time correspondence between $U_{in}$ and
$U_K$ is that the initial vacuum (containing no negative frequency
with respect to $u_{in}$ or $U_{in}$)
will rapidly evolve %(when $\kappa \neq 0$)
 into the Unruh vacuum
(the state containing no negative frequency with respect to $U_K$),
i.e. the transient flux will rapidly die out.
 The decay of these transients is governed
 by the difference between $U_K$ and $U_{in}$. Near the horizon,
  they are related by
 \be
 \label{UUK}
 U_{in} = D' U_K +   O(U_K^2),
 \ee
%R
where $D'$ is another irrelevant constant.
%R QUESTION 1 is the coef in the front of the cube FIXED by the geometry
%as it is the case of for EBH ?????????? There is a quadratic term U^2
%and both quadratic and cubic terms give a regular contibution on the horizon
%in the NEBH case
%IF so it shoudl be said
%if not it should be said as well
%The fact that they differ to cubic order only will
%play a crucial role in guaranteeing the regularity of the fluxes
%on the horizon.

We now consider a massless
minimally coupled scalar field propagating in the above collapsing geometry.
%above spacetime of the collapsing shell.
Taking the quantum state of the field
($|in\rangle$) to be Minkowski vacuum on ${\cal I}^-$ %inside the shell
implies that the state %of infa configurations
is vacuum with respect to the positive frequency modes \be
\phi_\omega^{in}(v) \propto e^{- i\omega v}, \quad
\phi_\omega^{in}(u) \propto e^{- i\omega U_{in}(u)} .
\label{inM}\ee As recalled in the Appendix, this determines the
expectation values of the stress tensor everywhere. Inside the
shell (i.e. $v<v_0$), we have \be \langle in| T_{\mu \nu
}|in\rangle\equiv 0 ,
%RS twice is too much \qquad v<v_0 ,
\ee
because the geometry is flat.
Outside the shell, in the BH geometry, the stress-tensor %RS fluxes
splits into a static part, which is %RS due to
completely determined by $f(r)=
1-{2M}/{r}+{Q^2}/{r^2}$ and which can be viewed as a vacuum
polarization, and a time-dependent outgoing flux  %RS part
which is caused by the
collapse: %(see the Appendix for details)
\bea
\langle in|
T_{uv}|in\rangle&=& %tr T_{VP} =
 -\frac{1}{24\pi}\bigg(1-\frac{2M}{r}+\frac{Q^2}{r^2}\bigg)
\bigg(\frac{M}{r^3}-\frac{3}{2}\frac{Q^2}{r^4}\bigg),
\label{Tuv}\\
\langle in| T_{vv}|in\rangle&=& %\langle in| T_{vv}(r)|in\rangle_{VP} =
\frac{1}{24\pi}\Bigg(-\frac{M}{r^3}
+\frac{3}{2}\frac{M^2+Q^2}{r^4}-\frac{3MQ^2}{r^5}+\frac{Q^4}{r^6}\Bigg),
\label{Tvv}\\
\langle in| T_{uu}|in\rangle&=&
%\frac{1}{24\pi}\Bigg(-\frac{M}{r^3} +\frac{3}{2}\frac{M^2+Q^2}{r^4}-\frac{3MQ^2}{r^5}+\frac{Q^4}{r^6}\Bigg)
\langle in| T_{vv}(r)|in\rangle
-\frac{1}{24\pi}\{u_{in},u\}(u).
\label{TuuH2}
\eea
The %RS only time-dependence comes from the
 outgoing flux is governed by the
Schwarzian derivative $\{u_{in},u\}$.
Using %evaluated form
Eqs. (\ref{uuinne},
\ref{newU})
we get
% gives

\bea
\{u_{in},u\}= \{U_{in},u\}&=&
%\nonumber\\&=&
-2\k_+^2\frac{\bigg[1-\frac{\k_+}{\k_-}\frac{U_{in}^3}
{(U_{in}-2(r_+ -r_-))^3}\bigg]}{\bigg[1-\k_+U_{in}
-\frac{\k_+}{\k_-}\frac{U_{in}}{(U_{in}-2(r_+ -r_-))}\bigg]^3}\nonumber\\
 &&+\frac{3}{2}\k_+^2\frac{\bigg[1-\frac{\k_+}{\k_-}
 \frac{U_{in}^2}{(U_{in}-2(r_+ -r_-))^2}\bigg]^2}
 {\bigg[1-\k_+U_{in}
 -\frac{\k_+}{\k_-}\frac{U_{in}}{(U_{in}-2(r_+ -r_-))}\bigg]^4}\ .\label{Suuinne}\eea
At early times, when the
shell radius is much larger than $r_+$,
$u\sim U_{in}\to -\infty$ and
%i.e.  for $u \ll v_0$,
the flux vanishes as one might expect.
%We now consider the
At late times, for $u \to+\infty$ and
 $U_{in} \to 0$,
 %of the above expression.% %and
% By a direct evaluation,
 we obtain
 \be\label{uuinnelimt}
\{U_{in},u\}\stackrel{u=+\infty}{\longrightarrow}-\frac{\k_+^2}{2}
+ C' U_{in}^2 .\ee
%R plesae check this is CRUCIAL (I think power 3  is correct, even though 2 is also
%acceptable from the point of view of regularity
%so what is the generic case 2 or 3 ??
%this is QUESTION 3
%and is C' ALSO indep of the details of the collapse, this is QUESTION 4
%do you know who has noticed 3,4 before ???
%
%%we cannot leave this unclarified because this term is central in the next section
%the generic case is power 2
The constant term %(Eq. (\ref{uuinnelimt}))
describes the stationary Hawking flux
at the temperature $T_H=\k_+/2\pi$.
It depends only on the final geometry, and is thus independent on the choice
of the collapsing configuration. Indeed, it coincides with the flux
%expectation value for the stress tensor
calculated in the Unruh vacuum, the stationary state where outgoing modes are
 positive frequency with respect to the Kruskal coordinate $U_K$
 of Eq. (\ref{KruskalUin2}). This directly follows
 from
 \be \label{schderK} \{U_K,u\} = -{\k_+^2}/{2}.\ee
 %and Eq. (\ref{uuinnelimt}).
From equation (\ref{uuinnelimt}), we also learn that the transient terms,
which depend on the details of the collapse,
%%R IS THIS CORRECT ??, is C' UNIVERSAL or COLLPASE DEPENDENT ?
%C' is collapse dependent
%and which
die out with two powers of $U_{in}$, i.e. like  $\exp(-2\k_+u)$ as
$u\to+\infty$ in terms of the asymptotic null time, and not only
with one power as one might have expected. In brief, at late
times, the outgoing flux becomes stationary and given by eq.
(\ref{TuuH2}) with $\{u_{in},u\}$ given by the first term of Eq.
(\ref{uuinnelimt}).
%
% \be\label{VPH} \langle T_{ab}\rangle=T_{ab}^{VP}+ T_{ab}^{H},\ee
% where the first term is the vacuum polarization part \bea
% T_{uu}^{VP}&=&\frac{1}{24\pi}\Bigg(-\frac{M}{r^3}
% +\frac{3}{2}\frac{M^2+Q^2}{r^4}-\frac{3MQ^2}{r^5}+\frac{Q^4}{r^6}\Bigg),\\
% T_{vv}^{VP}&=&T_{uu}^{VP},\\
% T_{uv}^{VP}&=&-\frac{1}{24\pi}\bigg(1-\frac{2M}{r}+\frac{Q^2}{r^2}\bigg)
% \bigg(\frac{M}{r^3}-\frac{3}{2}\frac{Q^2}{r^4}\bigg),\eea whereas
% the last term describes Hawking radiation \be\label{TuuH}
% T_{uu}^H=\frac{k_+^2}{48\pi}.\ee The other components of
% $T_{ab}^{H}$ vanish.\\
%It should be noticed that tThese expressions
%Eq. (15) coincides

% $u\to \infty$.
%Hence at late time lin> -> lUnruh >

The crucial property of the expectation values (\ref{Tuv}-\ref{TuuH2}) is their regularity
% of the stress tensor
on the future %R future is needed to expalin the diff between Tvv and Tuu
outer horizon. % $r=r_+$.
We remind the reader that regularity on the future horizon
%(defined in a freely falling frame)
requires that the energy density measured by a free falling observer
%RS $
\be
\rho_{FF}\equiv T_{\mu\nu}\frac{dx^\mu}{d\tau}\frac{dx^\nu}{d\tau}\, ,
\label{fFF}
\ee
 is finite. In the above $\tau$ is the proper time of the observer.
 In the limit $r \to r_+$,
%RS where
$\frac{dv}{d\tau}$ is constant and $\frac{du}{d\tau}\sim \frac{1}{f}$.
%RS This
The finiteness of $\rho_{FF}$ thus leads to the following conditions~\cite{Christensen:1977jc}
\bea
\lim_{r\to r_+}f^{-1} \, \langle T_{uv}\rangle &<& \infty,
\\
\lim_{r\to r_+}\langle T_{vv}\rangle &<& \infty, \\
 \lim_{r\to r_+}f^{-2}\, \langle
T_{uu}\rangle &<& \infty, \label{regularitya} \eea where
$f %=(1-2M/r+Q^2/r^2)
\to (r-r_+)(r_+-r_-)/r_+^2$. %R plaese check factors
The first two conditions are %R not trivially
satisfied since $\langle T_{uv}\rangle$ is state independent and vanishes linearly as
$r-r_+$ %R Plaese CHECK
 and since
$\langle T_{vv}\rangle$ is regular in the $in$ vacuum.
The last one requires more care.
%R I would propose to meke it in two steps
From Eqs. (\ref{nmnm}, \ref{fluxf},\ref{schderK})
%,\ref{kpm}) %VPH}) ($uu$ component)
we obtain  %have
that the late time limit (Unruh vacuum) behaves as %RS ($\kappa_+=f'(r_+)/2$)
%RS2 to be explicit
\bea \langle U \vert T_{uu} \vert U \rangle &=&
\langle in| T_{vv}|in\rangle + \frac{1}{48\pi} \kappa_+^2 ,
\nonumber \\&=&
-\frac{1}{192\pi}
\left( f'(r)^2 - 2f(r)f''(r) -f'(r_+)^2\right)  \, , \eea
where we have used $\kappa_+=f'(r_+)/2$.
In the
limit $r\to r_+$, we obtain %RS have that
%  - f'(r_+)f```(r_+)(r-r_+)^2$,
%thus
\be
\langle U \vert T_{uu} \vert U \rangle \stackrel{r\to
r^+}{\sim}
%\, C
%\nonumber\\
 f^2\,  C,\ee  where $ C=\frac{f'(r_+)f'''(r_+)r_+^4}{192\pi (r_+ - r_-)^2}$
is a constant depending on $M$ and $Q$.
% the mass and the charge of the black hole.
The steady part of the outgoing flux in the $in$ vacuum
is thus regular on the horizon \cite{Davies:1976ei}.

 It is equally important to notice that
 the transients which have been neglected above
 %in the stationary Unruh vacuum
 % in the late time limit
 do not spoil this regularity because they decrease with two powers of $r-r_+$
 as $r \to r_+$, see Eq. (\ref{uuinnelimt}). (We remind the reader
 that in a freely falling frame the following relations hold across
 the horizon:
 $d\tau \propto dU_{in}
 \propto - dr $ where $\tau$ is the proper time in this frame.)
 % shall we explain that in a FF frame dr propto dU
 %RS2 Toooo many 'Because'
These two powers compensate the divergence of $1/f^2 \sim 1/(r-r_+)^2$ in
eq. (\ref{regularitya}). Thus the transient contribution, taken alone, %RS2
is regular on the horizon.
%"The role of the transcients is simply
%%%to modify the above constant $C$ in a ((((non universal ????)))) manner.
%RS2 dear collaborators
%%could you please confirm the above point, that I raised in July,
%nor for the fun, but because it is directly RELEVANT
%%for eq 36.
%

 In conclusion, we have verified that the
% the first (relation (\ref{regularitya}))
regularity condition applied to the outgoing flux
is satisfied
%and the stress tensor is regular
on the outer future horizon of a NEBH both by the steady ``Unruh"
expectation values and by the late time transients.

\section{Hawking radiation emitted by EBH}

\subsection{The stationary expectation values}

Extremal BH have  are characterized by
$M=|Q|$ and their line element {\it can} %R seems obvious but is CRUCIAL for US
be obtained from Eq. (\ref{NEle}) by taking the limit $M \to Q$.
One gets \be \label{Ele}
 ds^2=-\bigg(1-\frac{M}{r}\bigg)^2dudv,\ee
 where $u=t-r^\ast$, $ v= t+r^\ast$ as before,
 and where the ``extremal" tortoise coordinate is
 \bea
 r^\ast(r;M) &=& \int^r\frac{dr'}{\Big(1-{M}/{r'}\Big)^2},
 \nonumber\\
  &=& r + M \Bigg[ \frac{- 1}{{r}/{M}-1} + 2 \ln({r}/{M}-1) \Bigg ] .
 \label{Erstar}\eea
 %R SQUARE was missing
 The novelty is that the merging of the two horizons
 causes a double zero of the metric $f= (1-{M}/{r})^2$
 on the horizon at $r=M$.
 As a consequence, the surface gravity vanishes, as can be seen
  by taking the limit $M \to Q$ in
Eq. (\ref{kpm}), and, equally important,
 $r^\ast$ now diverges as $-1/(r-M)$ when approaching the horizon
 and no longer as a logarithm in $r-r_+$ as was the
 case for NEBH in Eq. (\ref{NErstar}).

To obtain the %RS
stationary value of the
stress tensor for EBH, two approaches give the same result.
The first one consists in
performing the extremal limit $\k_+\to 0$ of the late time
limit of Eqs. (\ref{Tuv}-\ref{TuuH2}).
The second one consists in working directly
with the extremal metric given above,
%It should be noticed that one would have obtained the same expressions
% Eqs.(\ref{exwrong1}-\ref{exwrong2}-\ref{exwrong3}) would appear if one when
requiring that the stress tensor be % conserved,
static and vanish at infinity.
In the second approach, the trace anomaly determines unambiguously
the following %RS {\it stationary}  I do not think it is needed in insist here
expressions
%$\langle T_ {ab}\rangle_{EX}$ to be conserved, with the
%proper trace anomaly, static
% as a zero temperature
% obtained for non-extremal BH in the late time limit.
% Doing so one would obtain the following stress tensor
% for a non-extremal BH \be\label{Tabextwrong}\langle T_
% {ab}\rangle_{EX}= T_ {ab}^{VP}\bigg|_ {M=|Q|}.\ee Just the vacuum
% polarization term survives. In details
\bea
\langle T_
{uv}\rangle_{stat}&=&
-\frac{1}{24\pi}\frac{M}{r^3}\bigg(1-\frac{3}{2}\frac{M}{r}\bigg)
\bigg(1-\frac{M}{r}\bigg), \label{exwrong3}
\\
\langle T_{vv}\rangle_{stat}&=&-\frac{1}{24\pi}\frac{M}{r^3}\bigg(1-\frac{M}{r}\bigg)^3,
\label{exwrong2}\\
\langle
T_{uu}\rangle_{stat}&=&
\langle T_{vv}\rangle_{stat}.\label{exwrong1}
 \eea
%A common statement is  that in an extremal Reissner-Nordstr\"om BH
%the Boulware and Unruh (and the Hartle-Hawking) states coincide.
The %crucial point
novelty is that  Eq. (\ref{exwrong1}) does not vanish
sufficiently rapidly on the
horizon to fulfill the regularity condition.
In fact, since $f$ has a double zero, one gets
\be\label{divergence} \lim_{r\to M}f^{-2}\langle
T_{uu}\rangle_{stat}=\lim_{r\to
M}\left[-\frac{1}{24\pi}\frac{M}{r^3}\bigg(1-\frac{M}{r}\bigg)^
{-1}\right]=\infty.\ee
This implies that an observer free falling
across the horizon will measure an infinite energy density. % and pressure.
This fact, reinforced by the uniqueness of the stress tensor
%(\ref{exwrong1}-\ref{exwrong2}-\ref{exwrong3})
under the simple (and apparently sound)
%previous mentioned
hypothesis of stationarity and asymptotic vanishing flux \cite{Loranz:1995gc}
has led to the conjecture
%in the past to the believe
 that EBH might be singular objects from a quantum mechanical point
of view \cite{Trivedi:1992vh,Frolov:1987gw, Anderson:1994hg}. %RS Therefore they must
 In this case, they should
 be conceived as being
%The associated singularity was however considered to be "mild"
%since it gives rise to finite tidal distortion. In any case,
%extremal BH appeared to be completely dissociated
 disconnected from the regular NEBH (at least in two dimensions).

However in \cite{Balbinot:2004jx, Fagnocchi:2005uk} it was shown that by considering the
formation of a BH which is ab initio extremal, the resulting outgoing flux,
$\langle in \vert
T_{uu}\vert in \rangle$,
 %significantly in the BH region
radically differs from Eq. (\ref{exwrong1}).
Namely it is time dependent as one might have expected, but, more importantly,
its late time dependence is {\it universal}, i.e. independent
of the regular collapse one has chosen,
and such that
% the late time transcients are such that
$\langle in \vert T_{uu}\vert in \rangle$ is regular on the horizon.
One is therefore led to conclude that the above stationary
expressions $\langle
T_ {ab}\rangle_{stat}$ do not
%cannot be considered as characterizing
characterize the late time behaviour
of the stress energy of regular EBH.
% from the collapse of an object forming an extremal BH.
To see how regularity is achieved, let us
briefly review what happens when an EBH is formed by collapse.
%,see \cite{,,} for the original work.

\subsection{The flux emitted by an incipient EBH}
% Gerlach and Grove ? used this wording if i remember well, and I like it

 Consider the collapse of a charged null shell with $M=Q$.
% The line element reads \be
% ds^2=-\bigg(1-\frac{M}{r}\bigg)^2dudv,\ee where
% \bea &&u=t-r^\ast,\qquad v= t+r^\ast,\\
% && r^\ast=\int\frac{dr}{1-\frac{M}{r}}.\eea
In this case, repeating the steps of Section 2,
that is, using Eq.(\ref{Erstar}) in the place of Eq.(\ref{NErstar}), %{rstarshell
one finds
%at late time in the future of the shell instead of Eq. (\ref{KruskalU})
%(\ref{KruskalU})
 \bea \label{uuinex}
u
&=& u_{in}-4M\Bigg[
\frac{-1}{2\big (\frac{v_0-u_{in}}{2M}-1 \big )}
+ \ln \Big (\frac{v_0-u_{in}}{2M}-1\Big)
%\ln \Big (\frac{v_0-u_{in}}{2M}-1\Big)
%-\frac{1}{2\big (\frac{v_0-u_{in}}{2M}-1 \big )}
\Bigg]
\nonumber \\
 &=& U_{in} -4M\Bigg [
 %\ln \Big (- \frac{U_{in}}{2M}\Big) + \frac{M}{ U_{in}}
 \frac{M}{ U_{in}} + \ln \Big (- \frac{U_{in}}{2M}\Big)
 \Bigg ]
, \eea
%
% \be
% \label{uU}u=-4M\bigg[\ln\bigg(-\frac{U}{M}\bigg)+\frac{M}{2U}\bigg]
% ,\ee
where we have introduced the null coordinate
\be\label{uinext}U_{in} = u_{in} - v_0 + 2M\, ,\ee which again vanishes
on the future horizon.
%in this case $u_in=2U-2M+v_o$
%%%R ATTENCION To factors of 2 / MY U_in differs by a fcator of 2 from the old U
%and the horizon $r=M$ is at $U=0$
 Evaluating the
Schwarzian derivative $\{U_{in},u\}$ one gets
 \bea  \langle in|T_{uu}|in\rangle &=&
  \langle
T_{uu}\rangle_{stat} + \langle T_{uu}\rangle_{transients},
 \nonumber\\
 &=& \langle
T_{uu}\rangle_{stat}
 -\frac{1}{24\pi}\frac{8MU_{in}^3}{(U_{in}-2M)^6}
 %R \frac{U^3(U-2M)}{(2U-M)^4}
% \frac{U_{in}^3(U_{in}-4M)}{(U_{in}-M)^4}
 .\label{Tuuex}
%\\
% \langle in|T_{vv}|in\rangle&=&\langle T_{vv}\rangle_{EX}\label{Tvvex}\\
% \langle in|T_{uv}|in\rangle&=&\langle
% T_{uv}\rangle_{EX}\label{Tuvex}.
\eea %As for NEBH,
The last term %in Eq. (\ref{Tuuex})
describes the outgoing radiation. %RS2
At early times, for $U_{in} \to - \infty$, it
decreases with three powers of the affine null parameter $U_{in}$.
%RC is everything correct ?
At late times, for $u\to\infty$, it
%RS thus (is not needed since there is also a 'since' after)
vanishes \cite{Gao:2002kz}
as expected since there cannot be Hawking radiation,
the surface gravity being zero. In terms of the asymptotic time $u$
the transient flux vanishes as $1/u^3$, and not exponentially fast
as transients died out
%and no longer
 %(in the retarded time $u$) as it was the case
 for NEBH.
%%%%R is this RELEVANT ???
%%%%because regularity is given in terms of dr which is defines on the hor
%%%%so why the dep in u matters
%%%%I have reduced therefore the length of this paragraph, it is OK ??
%
% This behavior differs from that of NEBH in two respects.
% %should be compared with the corresponding one obtained
% %for non-extremal BH. In that case,
% First, for NEBH, at late time (see Eq. ) the
% outgoing radiation approaches the stationary thermal flux at the
% Hawking temperature up to transient terms decaying exponentially
% fast in the retarded time $u$.

Now let us examine the behavior of
Eq. (\ref{Tuuex}) %-\ref{Tvvex}-\ref{Tuvex})
when crossing the future horizon.
%The crucial equation is (\ref{Tuuex}), the regularity
%conditions for the $(vv)$ and $(uv)$ components are trivially satisfied.
We see that for $r\to M$ and $U_{in} \to 0$, the static vacuum
polarization term %RS3 $\langle T_ {uu}\rangle_{stat}$
and the transient  flux both vanish with three powers of $r-M \propto - U_{in}$.
%RS3 -U_{in}/2$. Rem. proto and the 1/2 is, to me and to some readers as well,
% meaningless.
%A here I inserted more details to show the cancellation of the
%divergent term. It makes the paper more self-contained and, also, avoids
%to refer to ref. 6 where the calculation was done for the late-time 'Unruh'
%and not, strictly speaking, for the in state
%RS2 good
Hence, if taken
 separately, both give a divergent contribution on the horizon, as
 shown in Eq. (\ref{divergence}).
 However, when expressing $U_{in}$ as a function of $r-M$ along an
 {\it arbitrary} infalling geodesic, i.e. with $dv/d\tau = \lambda$
 evaluated at the horizon characterizing the infalling velocity, one obtains
  %R without this how can one take their sum
 %%%unambiguously ?????
 %RS2 Sorry, this is 'troppo importante' to be left ambiguous
 %%Could you please write down how U_in depends on r near M
 %along an ARBITRARY infalling geodesic ?
 %(NOT only a light like one)
 %In fact, I do not see how this could work ?!
 %and I do not know what is the meaning of your '\sim'
 % In particular, I do not see why the normalisation of old36 and old37
 %% SHOULD be the SAME
 % Am I missing something ?
 % SO let be explicit:
 \be
 U_{in}(r) = -2 (r-M) + O((r-M)^2) \,  ,% ??? (r-M)^2 + O((r-M)^3) \,  ,
 \ee
 %RS3 do you know how lambda enters the second term ?
 where the first term is independent of $\lambda$. (This follows from
 the light-like character of $dr$ on the horizon).
 %RS3 you see, the trouble I had is that
 % in flat space the Jacobian du/dr along a geodesic IS a function of lambda
 This independence guarantees that the leading terms of the
 two contributions of eq. (\ref{Tuuex}) cancel each other:
 \bea
  \langle in|T_{uu}|in\rangle %RS2 \sim
   &=&    \langle
T_{uu}\rangle_{stat} + \langle T_{uu}\rangle_{transients}\, ,
  \nonumber\\
  &=&   -\frac{(r-M)^3}{24\pi M^5} -\frac{8MU_{in}^3}{24\pi (2M)^6}
  +O((r-M)^4)\, ,\nonumber\\
  &=&  O((r-M)^4)\, .
  \eea
%  since \be \langle
% T_{uu}\rangle_{stat} %RS2 \sim
% = -\frac{(r-M)^3}{24\pi M^5} + O((r-M)^4), \ee and \bea
% \langle T_{uu}\rangle_{transients}  %RS2 \sim
% &=& -\frac{8MU_{in}^3}{24\pi
% (2M)^6}+ O(U_{in}^4), \nonumber\\
% &=&  \frac{(r-M)^3}{24\pi M^5} + O((r-M)^4) ,\eea
% implying that \be
%   \langle in|T_{uu}|in\rangle %RS2 \sim
%   =
%  O((r-M)^4) , \ee and so
Hence $f^{-2}\langle in|T_ {uu}|in\rangle $ stays finite.
% Indeed a direct evaluation gives\cite{wherewasthisfirstshown?}
% \be\lim_{r\to
%M}f^{-2}\langle in \vert
%T_{uu} \vert in \rangle=-\frac{1}{24\pi}\bigg(\frac{15}{2M^2}\bigg)<\infty.\ee
%R QUESTION 4
%is this constant UNIVERSAL or collapse dep ??
% we shoudl say a word on this
The key point is that, even though the logarithmic term in Eq. (\ref{uuinex})
is {\it subleading} at late times,
%RS3 not necesseary given the above details for $U_{in} \to 0^{-}$,
%RS2 the presence of
this term
is necessary to get the above cancellation.
%RS3 a finite result. %this final result, .
Indeed, its omission would give a vanishing
Schwarzian derivative %$\{U,u\}$
and therefore would give
back the singular behaviour of $\langle T_ {uu}\rangle_{stat}$ %RS .
obtained in the former subsection. Notice that this term was
omitted in \cite{Liberati:2000sq}, see eq. (3.4),
thereby leading
%R3 This led the authors
to the erroneous conclusion that there is a ``real
discontinuity" between NEBH and EBH.

In brief, for EBH formed by the collapse of a shell, or more
generally formed by a regular collapse \cite{Fagnocchi:2005uk},
the stress tensor is in fact regular on the horizon.
Having reached this conclusion, we finally arive at the %crucial
question we wanted to confront:
 why was this regular behavior missed
 in the %analysis of the $\k_+\to 0$
extremal limit giving rise to
%of the collapse of a non-extremal BH which, as we have seen
Eq. (\ref{exwrong1}) ?
%, leads to the singular $\langle T_
%{ab}\rangle_{EX}$?

\subsection{The extremal limit of the non extremal flux}

To answer the above question,
%Before reaching the final conclusion, a more careful analysis
one should reconsider how to implement the extremal limit. % $\k_+\to 0$.
%he above extremal limit has been taken is compulsory. One should
To this end, we first note that
%in the non-extremal BH collapse,
although the late time value of the Schwarzian derivative
$\{U_{in},u\}$  has a smooth limit for $\k_+\to 0$,  as shown in
Eqs. (\ref{Suuinne},\ref{uuinnelimt}), thereby leading to the
singular result of Eq. (\ref{divergence}),
%$\langle T_ {ab}\rangle_{EX}$ and to all the related analysis,
the late time behaviour of $u(U_{in})$, which is that of
Eq.  (\ref{KruskalUin2}),
% between
%the Kruskal coordinate
%$U_{in}$ and $u$
has no well defined limit $\k_+\to 0$.
% % Remember
% this relation has been obtained from the matching condition and
% performing the late time limit ($u\to\infty$, $U\to 0$). This limit
% ha to be taken with care if one wants at the end examine extremal
% configurations, since for these only transient terms appear and the
% way they vanish is crucial for the regularity issue. So let us come
The ill-defined character of this extremal limit % of Eq. (\ref{KruskalUin2})
tells us that the late time limit should not have been taken first.
%before the extremal limit.

So let us return to Eq. (\ref{uuinne}) which gives
the exact relation between $U_{in}$ and $u$
%\emph{out} modes in the collapse leading to a non-extremal BH.
and %We
perform the extremal limit first. We notice that
in terms of the surface gravities this limit reads $\k_+\to 0$,
$\k_-\to 0$, $\k_+/\k_-\to 1$. So instead of using these as parameters,
we shall re-express the expressions directly in terms of $M$ and $Q$,
and simply send $M \to Q$.
%.\\in this expression, instead of the late time limit as done before, directly
We also notice that the non-trivial character of this limit
%applied to Eq. (\ref{uuinne})
% the function $u(U_{in};M,Q)$
entirely comes through the tortoise coordinate $r^\ast(r;Q,M)$
which enters in the second equation of Eqs. (\ref{rstarshell}).
It is therefore {\it sufficient}
to study the extremal limit of $r^\ast(r;Q,M)$ of Eq. (\ref{NErstar}).

So, working at fixed $r$ (which amounts to not taking the late
time limit), forming the half difference and the half sum of the
two logarithms to sort out the singular and regular part, then
using the values of $\k_\pm$ and $r_\pm$ of Eqs. (\ref{kpm}), and
only then performing the limit $M \to Q$, we successively get
  %Starting from the second of Eqs. (\ref{rstarshell})
  \bea
%u &=& -2r -\frac{2M^2-Q^2}{\sqrt{M^2-Q^2}}\ln \bigg |\frac{r-r_+}{r-r_-}\bigg |-2M\ln |(r-r_+)(r-r_-)| + v_o\nonumber\\
2 r^\ast(r; Q,M) -2r &=&
%\frac{2M^2-Q^2}{\sqrt{M^2-Q^2}}\ln \bigg |\frac{r-r_+}{r-r_-}\bigg | + 2M\ln %|(r-r_+)(r-r_-)|
\Bigg ( \frac{1}{2\k_+} + \frac{1}{2\k_-}  \Bigg )
\ln \bigg |\frac{r-r_+}{r-r_-}\bigg | \nonumber\\ && +
\Bigg ( \frac{1}{2\k_+} - \frac{1}{2\k_-} \Bigg ) \ln |\frac{(r-r_+)(r-r_-)}{r_+r_-}|
, \nonumber\\
% &=&  -2r -\frac{2M^2-Q^2}{\sqrt{M^2-Q^2}}\ln \bigg |\frac{r-(M+\sqrt{M^2-Q^2})}{r-(M-\sqrt{M^2-Q^2})}\bigg |
% &&-2M\ln |(r-r_+)(r-r_-)| + v_o\nonumber\\
&=&  \frac{2M^2-Q^2}{\sqrt{M^2-Q^2}}\ln \bigg |1-\frac{2\sqrt{M^2-Q^2}}{r-M+\sqrt{M^2-Q^2}}\bigg |
\nonumber\\ &&+ 2Q\ln |\frac{(r-r_+)(r-r_-)}{r_+r_-}|, %+ v_o
\nonumber\\
&\stackrel{M\to Q}{\longrightarrow}&   % -2r -
 \frac{2M^2-Q^2}{\sqrt{M^2-Q^2}}\times
\frac{-2\sqrt{M^2-Q^2}}{r-M+\sqrt{M^2-Q^2}} + 4Q\ln \frac{(r-Q)}{Q},
% -2M\ln |(r-r_+)(r-r_-)| +v_o
\nonumber\\
%eea \be u
&\stackrel{M\to Q}{\longrightarrow}&
%-2r+
- \frac{2Q^2}{r-Q} + 4Q\ln \frac{(r-Q)}{Q}, %+v_o. \ee
%\nonumber
\label{Elimr}
\eea
thereby recovering the behavior of the
extremal tortoise coordinate one obtains from the extremal meric
in eq. (\ref{Erstar}).
Therefore, when eliminating $r$ using the first of Eqs. (\ref{rstarshell}) we
also recover eq. (\ref{uuinex}) which gives rise to a regular flux.
%
% eventually arives
%  \be \label{uuinex}
% u=u_{in}-4M\Bigg [\ln \Big (\frac{v_0-u_{in}}{2M}-1
% % \Big)-\frac{1}{2\big (\frac{v_0-u_{in}}{2M}-1 \big )} \Bigg ]. \ee
% %
% What one finds is exactly the relation relating the \emph{in} and
% the \emph{out} modes one obtains for the collapse of an extremal
% black holes (see Eq. (\ref{uU})).
%Therefore, we have

We have thus established that the extremal %$\kappa_+\to 0$
limit  $M \to Q$ of the exact relation (\ref{uuinne})
smoothly connects to the extremal expression $u(U_{in};Q)$
of eq. (\ref{uuinex}). In addition, since the extremal limit
applied to the late time expression of (\ref{uuinne})
given  %$u(U_{in};M,Q)$,
in Eq.  (\ref{KruskalUin2}) is ill-defined,
we %One concludes
have demonstated that the late time limit (i.e. the stationary limit)
and
the extremal %$\kappa_+\to 0$
limit % clearly
do not commute.

This non-commuting character explains why the
fluxes obtained using the extremal limit of the exact
relation are regular even though those obtained by taking
the extremal limit of the
stationary fluxes were singular on the horizon.
%We leave as an exercice for the interesting reader
%to verify that %complete the proof by verifying that
%To complete the analysis,
% If one can consider
Indeed, it is not difficult to show that the extremal limit of the
Schwarzian derivative governing the collapse of a NEBH given in
Eq. (\ref{Suuinne})
leads %RS indeed too many indeeds
to the transients present in Eq. (\ref{Tuuex})
%late time
which are necessary to preserve the regularity on the horizon.
% one gets for the extremal black hole collapse, see
% that is, which behaves at late time
 %RS  as it was the case and which are .
At fixed $u_{in}$ and in terms of the parameter
$\Delta\equiv\sqrt{M^2-Q^2}$,
 when using the coordinate $U_{in}$ of eq. (\ref{uinext}) to
simplify the expression, %RS
the Schwarzian derivative of
Eq. (\ref{Suuinne}) %RS Tuuex}) PLEASE CHECK
%is rewritten as
reads
\bea
%\{u_{in},u\}= &=&
%\nonumber\\&=&
% -\frac{2\Delta^2}{(M+\Delta)^2}\frac{\bigg[1-\frac{(M-\Delta)^2}{(M+\Delta)^2}
% \frac{(u_{in}-v_0+2M+2\Delta)^3}
% {(u_{in}-v_0+2M-2\Delta)^3}\bigg]}{\bigg[1-\frac{\Delta}{(M+\Delta)}(u_{in}-v_0+2M+2\Delta)
% -\frac{(M-\Delta)^2}{(M+\Delta)^2}\frac{(u_{in}-v_0+2M+2\Delta)}{(u_{in}-v_0+2M-2\Delta)}\bigg]^3}&&\nonumber\\
% +\frac{3\Delta^2}{2(M+\Delta)^2}\frac{\bigg[1-\frac{(M-\Delta)^2}{(M+\Delta)^2}
%  \frac{(u_{in}-v_0+2M+2\Delta)^2}{(u_{in}-v_0+2M-2\Delta)^2}\bigg]}
%  {\bigg[1-\frac{\Delta}{(M+\Delta)}(u_{in}-v_0+2M+2\Delta)
%  -\frac{(M-\Delta)^2}{(M+\Delta)^2}\frac{(u_{in}-v_0+2M+2\Delta)}
%  {(u_{in}-v_0+2M-2\Delta)}\bigg]^4}\, .
%
%
  -\frac{2\Delta^2}{(M+\Delta)^2}\frac{\bigg[1-\frac{(M-\Delta)^2}{(M+\Delta)^2}
\frac{(U_{in}+2\Delta)^3} {(U_{in}-2\Delta)^3}\bigg]}
{\bigg[1-\frac{\Delta}{(M+\Delta)^2}(U_{in}+2\Delta)
-\frac{(M-\Delta)^2}{(M+\Delta)^2}\frac{(U_{in}+2\Delta)}
{(U_{in}-2\Delta)}\bigg]^3}&&\nonumber\\
+\frac{3\Delta^2}{2(M+\Delta)^2}\frac{\bigg[1-\frac{(M-\Delta)^2}{(M+\Delta)^2}
 \frac{(U_{in}+2\Delta)^2}{(U_{in}-2\Delta)^2}\bigg]^2}
 {\bigg[1-\frac{\Delta}{(M+\Delta)^2}(U_{in}+2\Delta)
 -\frac{(M-\Delta)^2}{(M+\Delta)^2}\frac{(U_{in}+2\Delta)}
 {(U_{in}-2\Delta)}\bigg]^4}\, .
  && \label{Suuinneuin}\eea
%RS Expanding in terms of small $\Delta$ and t
Taking the extremal limit
$\Delta \to 0$ we are left with
\be \frac{8M U_{in}^3}{(U_{in}- 2M)^6} \, ,\ee
which  exactly %by use of Eq. (\ref{uinext}) gives
gives the transients of eq. %RS late time transients in the extremal flux
(\ref{Tuuex}). %RS already said necessary to ensure regularity on the horizon.

%and explicitely show that one gets, as one should,
%the extremal limit one obtains, with no surprise now, exactly
%a result which

 %Even though it is instructive to see explicitly how the
 %various terms combine to preserve the regularity,
 As in Eq. (\ref{Elimr}),
  the proof relies on
%  follows from
  the fact that the function $r^\ast(r;Q,M)$ of eq. (\ref{NErstar})
  {\it uniformly} converges to $r^\ast(r;M)$ of eq. (\ref{Erstar}) in the
  limit $M \to Q$, outside the horizon, since the integrand of $r^\ast(r;Q,M)$
  ($=1/f(r;M,Q)$)  is a differentiable function of $M$ and $Q$.
  %%Do you agree ??
  %%%R QUESTION 5 do you remember
  %%%%what are (is) the SUFFICIENT conditions for a Riemanian integral
  % to guarantee the uniformity of the convergence ????
  %in particular CAN "  outside the horizon" be omitted ? or not ?
  % I think it is appropriate here to be more rigorous and abstract mathematiaclly
  %than just verifying things explicitely
  %THE usefulness is that the math theorem should also work
  %for higher order functions and not only <Tmunu> !

%
%
% \bea
% \{u_{in},u\}&=&-2k_+^2\frac{\bigg[1-\frac{k_+}{k_-}\frac{(v_0-u_{in}-2r_+)^3}{(v_0-u_{in}-2r_-)^3}\bigg]}
% {\bigg[1+k_+(v_0-u_{in}-2r_+)-\frac{k_+}{k_-}\frac{(v_0-u_{in}-2r_+)}{(v_0-u_{in}-2r_-)}\bigg]^3}\nonumber\\
% &&+\frac{3}{2}k_+^2\frac{\bigg[1-\frac{k_+}{k_-}\frac{(v_0-u_{in}-2r_+)^2}{(v_0-u_{in}-2r_-)^2}\bigg]^2}
% {\bigg[1+k_+(v_0-u_{in}-2r_+)-\frac{k_+}{k_-}\frac{(v_0-u_{in}-2r_+)}{(v_0-u_{in}-2r_-)}\bigg]^4}.\nonumber
% \eea \be \{u_{in},u\}\stackrel{M\to
% Q}{\longrightarrow}-\frac{8Q(v_0-u_{in}-2Q)^3}{(v_0-u_{in})^6}=\{u_{in}^{EX},u\}.
% \ee \be u_{in}=2U-2M+v_0\ee
%

\section{Conclusions}

We have seen how a superficial way of treating the extremal limit %$k_+\to0$ limit
leads to the conclusion that the stress tensor of %extremal
EBH is singular %have pathological features
on the horizon, unlike what is found for NEBH. %in non extremal ones.
We have also shown that the singular behaviour
%This belief, commonly accepted in the literature,
results from having assumed the stationarity of the fluxes,
which amounts to neglecting
%in the extremal case
transients which are necessary for insuring the regularity on the horizon.

With more details,
when starting from the late time expressions of the non-extremal case,
one simply misses these transients because
%contribution which in this limit is indeed negligible for non extremal
they are negligible far away from the hole
when compared to the finite Hawking radiation, and on the horizon
they only give a finite and regular contribution comparable with that
of the steady part that decreases as $(r-r_+)^2$.
%which does not affect the
%because they decrease like $(r-r_+)^3$
%whereas the steady part decreases only like $(r-r_+)^2$
%which is sufficient to insure regularity for NEBH.
%%R is all this Correct ??
%when compared to the finite Hawking radiation
However, for EBH, because of the double zero
of the metric function $f(r)$ on the horizon,
regularity now requires that the outgoing part of stress tensor
vanishes with four powers of $r-M$. This, toghether with the fact that
 the steady part only vanishes with three powers, explains
 why the transients are not only necessary to preserve the regularity:
 they must be such that when combined with the steady part,
 the sum vanishes with four powers. Hence they must vanish with three
 powers and with a normalization which is independent of the collapse.

We have also demonstrated that when starting with the fluxes of NEBH
% but on the other hand is the only surviving term for
% extremal BHs since for these latter Hawking radiation is absent. The
and applying the
%correct limiting procedure is, as we have seen, to keep all terms and perform the
extremal limit ($M\to Q$) before the late time limit, the regularity
is preserved including in the limit, the EBH case. %extremal case.
% In this way
With this we %RS have proven
establish that EBH
should not %RS no longer
 be considered as pathological,
since their fluxes are smoothly connected to those
of NEBH.
%a smooth way to non extremal configurations.
Perhaps the most unexpected result is that
the transients fluxes that were negligible at late times
for NEBH evolve (as $M \to Q$) into the necessary transients
which cancel out the divergence of the static energy density
on the horizon.
What is also unexpected is that their late time behavior
%RS of these transients
is
independent of the (regular) collapse one is dealing with:
Explicitely, the second term of eq. (\ref{Tuuex}) behaves as
%R to refine this we must have firts clarify whther
%their universal charcater was also present for NEBH
%or ONLY appears in the extremal limit
\be \lim_{U_{in}\to 0} \langle
T_{uu}\rangle_{transients} = \alpha U_{in}^3 + O(U_{in}^4)\ ,\ee
where \be
\alpha=-\frac{1}{192\pi M^5}\ ,\ee
is indeed %RS model
collapse independent. We have also shown that the regularity of
the quantum expectation values follows from the well-defined
character of the function  $r^\ast(r;Q,M)$ of eq. (\ref{NErstar})
in the extremal limit. This is not suprising since the geometrical
optics approximation is exact in two dimensions, i.e., the
positive frequency $in$ modes of eq. (\ref{inM}) are entirely
governed by the classical function $U_{in}(u;Q,M)$. Therefore the
quantum expectation values can only depend on this function and
its derivatives (and possibly also on the local metric function
$f(r;M,Q)= dr/dr^\ast$). Since $r^\ast(r;Q,M)$ is $C^\infty$ in
$M,Q$, so are $U_{in}(u;Q,M)$ and its derivatives.

Finally we discuss the relevance of our conclusions
to four dimensional %RS
(or higher than bidimensional)
 black holes.
The fact that for EBH the transients are singular and cancel out the
divergence of the static energy density should also be found
in any dimension in spite of the presence of "grey-body" factors resulting from
the elastic scattering on the static centrifugal barrier.
Indeed, the Bogoliubov transformation relating, at fixed angular momentum,
 %in the geometrical optic approximation
the regular $in$ modes (\ref{inM}) to the positive frequency
$out$ modes $e^{-i\omega u}$
should possess properties which are independent of the
dimensionality, because the latter are singular on the horizon.
%in {\it any} dimension,
It is therefore difficult to conceive that the %RS elastic scattering
value of ``grey-body" factor
could interfere with %RS this transformation
the Bogoliubov coefficients in such a way as to
 give rise to transients which are regular on the horizon.
In fact, when assuming that the stress tensor obtained by considering
 a regular collapse be regular in any dimension,
 this leaves only two possibilities:
 either the divergence of the late time transients
 cancels out that of the static energy density (as it is the case in 2D \cite{Trivedi:1992vh}
and as found in Refs. \cite{Frolov:1987gw, Anderson:1994hg}),
  or they are both regular. Our reasoning concerning the modes at fixed
  angular momentum suggests that it is unlikely that the second option be
 realized. However, this contrasts with the numerical
 analysis of \cite{Anderson:1995fw} (see also \cite{Carlson:2003ub} for spin 1/2 fields) which concluded that the static energy density
 is regular for 4D EBH. We are planning to report on this with more details
 in a future %RS another
 paper.

%%R I think we should see/establish if this WAS also the case for NEBH
%in order to be able to comment on it §§

%%%R I think I do not agree with the old text below
% that for the collapse leading to the formation of
% a non extremal BH a stationary state is reached asymptotically in
% time (the so called Unruh vacuum) whereas no stationary state is
% approached in the formation of an extremal BH, the resulting state
% is unavoidably time dependent. This is not a surprise given the
% absence of Hawking radiation.

\section*{Acknowledgements}

\noindent  We thank P. Anderson and S. Liberati for interesting
comments. A. F. acknowledges the Spanish grant
FIS2005-05736-C03-03 and the EU Network MRTN-CT-2004-005104 for
financial support.

%\\
\appendix
\section{2D stress tensor}
%\section{Appendix}
%\\

In this Appendix we present the basic properties of the
stress energy tensor of a 2D massless field propagating
in a stationary metric, since this is all we need in the
body of the paper.

%Let us
We thus consider %a
2D spacetimes which %, for the purpose of the paper, is
are static and described by the metric \be\label{medonu}
ds^2=-f(r)dt^2+f^{-1}(r)dr^2 .\ee Introducing the null coordinates
%\bea
$u=t-r^\ast$, $v=t+r^\ast $
 where \be r^\ast=\int\frac{dr}{f(r)}\ee
the metrix is conformally flat  \be ds^2=-f(r)\,
dudv.\label{2Dnull}\ee
Therefore, a massless minimally coupled %quantum
scalar field satisfying the d'Alembert equation
will obey,  %, $\Box\phi=0$.
in double null coordinate system, the simplified equation
%. where
% %$\Box=\nabla^\mu\nabla_\mu$ is the covariant D'Alembertian.\\
%When the metric is expressed in a double null form, like Eq. (\ref{2Dnull}), the field equation
\be
\partial_u\partial_v\phi=0 .\ee
Its general solution is thus a sum of a function of only $u$ and
one of $v$.

Expanding the field operator in the positive frequency plane
waves $e^{-i\omega u}$, $e^{-i\omega v}$ defines a vacuum
state, say $|u,v\rangle$, by annihilation with the destruction operators
associated with these positive frequency modes.
Two basic properties of the renormalized expectation values of
the quantum stress tensor of $\phi$ are used in the text.

First, when  $\phi$ propagates in the space-time described by
(\ref{2Dnull}), its stress tensor reads \bea \langle u,v|
T_{uu}|u,v\rangle&=&\langle u,v|
T_{vv}|u,v\rangle=-\frac{1}{192\pi}(f'^2-2ff''),\label{fluxf}\\
\langle u,v| T_{uv}|u,v\rangle&=&\frac{1}{96\pi}ff'',\eea where a
prime indicates derivative with respect to $r$.

The second property follows from the fact that the set of
positive frequency modes
%built with the modes
$(e^{-i\omega u}$, $e^{-i\omega v})$
is not unique, even though it is complete.
One could introduce two new null coordinates
\be
U= U(u),\quad V=V(v),
\ee
and use these to define a new set of positive frequency modes
($e^{-i\lambda U}$,
$e^{-i\lambda V}$).  By the same procedure as above, these modes
%as basis for the expansion of the field operator $\phi$,
%where $\bar{u}=\bar{u}(u)$ and
%$\bar{v}=\bar{v}(v)$. This alternative expansion defines
can be used to define another vacuum state, named $|U,V\rangle$.
Then the expectation
values of the stress tensor in this new state are
related to the former one by
 \bea \langle U,V| T_{uu}|U,V \rangle&=&\langle u,v|
T_{uu}|u,v\rangle-\frac{1}{24\pi}\{U,u\}\label{nmnm}\\
\langle U,V | T_{vv}|U,V \rangle&=&\langle u,v|
T_{vv}|u,v\rangle-\frac{1}{24\pi}\{V,v\}\\
\langle U,V | T_{uv}|U,V \rangle&=&\langle
u,v| T_{uv}|u,v\rangle,\label{traceterm}\eea where $\{U,u\}$
is %ndicates
the Schwarzian derivative \be
\{U,u\}=\bigg(\frac{d U}{du}\bigg)^{-1}\frac{d^3 U}{du^3}
-\frac{3}{2}\bigg(\frac{dU}{du}\bigg)^{-2}(\frac{d^2U}{du^2})^2,\ee
and similarly for $\{V,v\}$. Eq. (\ref{traceterm}) is a
consequence of the state independence of the trace anomaly.

\end{document}